\documentstyle[aps,graphicx,floats,prb]{revtex}

\def\u2{$\langle u^{2} \rangle$}

\begin{document}
\draft
\wideabs{

\title{Measurement of the local Jahn-Teller distortion in LaMnO$_{3.006}$}

\author { Th. Proffen, R. G. DiFrancesco, S. J. L. Billinge } 
\address{ Department of Physics and Astronomy and Center for
          Fundamental Materials Research,\\
          Michigan State University, East Lansing, Michigan 48824-1116.}
\author { E. L. Brosha and G. H. Kwei}
\address{ Los Alamos National Laboratory, Los Alamos, New Mexico 87545.} 

\date{March, 4, 1999}

\maketitle


\begin{abstract}
The atomic pair distribution function (PDF) of stoichiometric LaMnO$_3$
has been measured.  This has been fit with a structural model to extract
the {\it local} Jahn-Teller distortion for an ideal Mn$^{3+}$O$_6$
octahedron.  These results are compared to Rietveld refinements of
the same data which give the {\it average} structure. Since the 
{\it local} structure is being measured in the PDF there is no
assumption of long-range orbital order and the real, local, Jahn-Teller
distortion is measured directly.  We find good agreement both with published 
crystallographic results and our own Rietveld refinements 
suggesting that in an accurately stoichiometric material there is long 
range orbital order as expected. The local Jahn-Teller distortion has 
2 short, 2 medium and 2 long bonds.  This implies that there is some 
mixing of the $d_{3z^2-r^2}$ and $d_{x^2-y^2}$ states and the occupied 
state is not pure $d_{3z^2-r^2}$ symmetry.
The Debye temperature of the Mn and O ions has also been calculated
as $\Theta_{D}$(Mn)$= 1000 \pm 100$~K, $\Theta_{D}$(O$_{\mbox{apical}}$)$= 
980 \pm 30$~K and $\Theta_{D}$(O$_{\mbox{basal}}$)
$= 601 \pm 8$~K. 
\end{abstract}

}


\section{Introduction}

The Jahn-Teller (JT) distortion of the MnO$_6$ octahedra 
in perovskite manganites is known to have a significant effect 
on their electrical and magnetic properties. The JT distortion
takes the form of an elongation of the octahedra.  The simplest type
of distortion is one in which the octahedra elongate along the
direction of the $d_{3z^2-r^2}$ orbitals and contract in directions 
in which the $d_{x^2-y^2}$ orbitals point.~\cite{griff;b;64}  This 
distortion would give rise to 2 long Mn-O bonds and 4 short 
Mn-O bonds in each distorted octahedron.
However, it is possible to generate a symmetry lowering distortion with
a different symmetry by making some linear combination of the pure
$d_{3z^2-r^2}$ and $d_{x^2-y^2}$ states.  Because of the importance of
the Jahn-Teller distortion in these materials it is critical to
characterize the exact nature of the JT state in the manganites.

The MnO$_6$ octahedra pack together in space in a 3-dimensional corner
shared network giving rise to the well-known perovskite structure.
In general the distorted octahedra can be orientationally ordered 
(so-called orbital ordering) or disordered.  If the orbitals are long-range
ordered then a solution of the average crystal structure, as obtained
from Rietveld refinement for example, will give
the local bond-lengths in the octahedra accurately and reveal the nature
of the local JT distortion.  However, if the orbitals are not perfectly
long-range ordered, the average crystal structure will not give the right
result for the local JT distortion.  However, a local structural probe
such as extended x-ray absorption fine structure (XAFS) or the 
atomic pair distribution function (PDF) method, will still reveal the
nature of the local distortion regardless of whether the orbitals are 
ordered or not.  

Any determination of the nature of the local JT distortion using a 
crystallographic approach necessarily presumes perfect orbital order.  
This is thought to be good in the case of undoped LaMnO$_3$ where every
manganese ion is in the 3+ state.  However, by measuring the local
structure directly using the PDF method, we do not make this presumption.
We have measured the local JT distortion in a sample of composition
LaMnO$_{3.006}$ using the PDF analysis of neutron powder diffraction data.
The PDFs we measure are essentially sample, and not resolution, limited 
and the short and long bonds in the distorted MnO$_6$ octahedra are 
clearly resolved. These PDFs have been modelled using a full-profile 
least-squares refinement approach. These results are compared to 
crystallographic Rietveld refinements on the
same data.  We find excellent agreement for the Jahn-Teller distortion 
between the PDF and crystallographic analyses.

The average crystal structure of undoped LaMnO$_3$ has been extensively
studied since the  1950's.\cite{wold;jpcs59,wolla;pr55,elema;jssc71,%
tofie;jssc74,roosm;jssc91,roosm;jssc94i,norby;jssc95,mitch;prb96,%
huang;prb97,rodri;prb98} Differences between these studies occur largely 
because of the sensitivity of the structure to the sample stoichiometry 
which depends on synthesis conditions.\cite{mitch;prb96}  
It appears fairly widely accepted now that the correct structure for 
stoichiometric LaMnO$_3$ at low temperature is orthorhombic
(space group $Pbnm$ or $Pnma$ depending on convention).  The data assigned
to a monoclinic space group by Mitchell {\it et al.}\cite{mitch;prb96}
can be well refined in the orthorhombic space group as well with fewer
degrees of freedom.\cite{rodri;prb98}  An excellent summary of the situation
is presented in Rodriguez-Carvajal {\it et al.}\cite{rodri;prb98}  In
this structure the long $d_{3z^2-r^2}$ orbitals lie in the same (basal) plane
in a checkerboard type of arrangement so the bonds are long-short-long-short
as you move from Mn to Mn along the Mn-O-Mn bond. Since all the long bonds
lie in this plane, the separation of the Mn ions in the 
perpendicular direction ($c$-axis in the $Pbnm$ setting
and $b$ axis in the $Pnma$ setting) is shorter.  This is the O$^\prime$ 
structure in the Goodenough specification.\cite{goode;pr55} 

There is one report of a PDF measurement on the undoped LaMnO$_3$ 
material.\cite{louca;prb97} In this case the monoclinic structure of 
Mitchell~{\it et al.}\cite{mitch;prb96} was 
successfully fit to the data.  However, no structural parameters were 
published. In this paper we publish the local structure parameters
of LaMnO$_3$ determined from PDF data.
The results are compared to a Rietveld refinement of the same data-set.
There is excellent agreement between the average and the local structures
indicating that the sample is fully long-range ordered.  
We find that, even locally, there is a significant orthorhombic distortion
to the MnO$_6$ octahedra.


\section{Experimental}

The LaMnO$_{3+\delta}$ sample was prepared using standard solid state reaction
methods. Stoichiometric amounts of La$_2$O$_3$ (Alfa Aesar Reacton 99.99\%) and
MnO$_2$ (Alfa Aesar Puratronic 99.999\%) were ground in an Al$_2$O$_3$ 
mortar and pestle under acetone until well mixed. The powder sample was 
loaded into a 3/4" diameter die and uniaxially pressed at 1000~lbs. 
The pellet was placed into an Al$_2$O$_3$ boat and fired under pure 
oxygen for 12 hours at 1200-1250~$^\circ$C. The sample was cooled to 
800~$^\circ$C and removed, reground, repelletized,
and refired at 1200-1250~$^\circ$C for an additional 24 hours. This process
was repeated until a single phase, rhombohedral, x-ray diffraction
pattern was obtained. Total reaction time was approximately 5 days. 
Thermogravimetric analysis (TGA) indicated that the
as-prepared sample had an oxygen stoichiometry of about LaMnO$_{3.10}$. 

The LaMnO$_{3.10}$ sample was ground, left in powder form, and placed into an
Al$_2$O$_3$ boat. The sample was post-annealed in ultra high purity
 Ar at 1000~$^\circ$C for 24 hours then quenched to room temperature. 
The oxygen stoichiometry was again determined using TGA
under forming gas. The final oxygen stoichiometry was 3.006.

\begin{figure}[!tb]
  \centering
  \includegraphics[angle=0,width=2.8in]{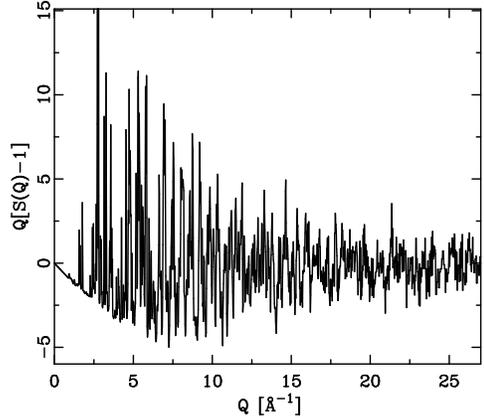}
  \caption{Reduced structure factor $F(Q)=Q[S(Q)-1]$ for LaMnO$_{3}$ measured
           at 10~K. The maximum value of $F(Q)$ is 26.} 
  \label{fig;soq}
\end{figure}

Neutron powder diffraction data were collected on the SEPD diffractometer
at the Intense Pulsed Neutron Source (IPNS) at Argonne National Laboratory.
The sample of about 10~g was sealed in a cylindrical vanadium tube with helium 
exchange gas. Data were collected at 20K in a closed cycle helium refrigerator.
The data are corrected for detector deadtime and efficiency, 
background, absorption, multiple scattering, inelasticity effects and
normalized by the incident flux and the total sample scattering cross-section 
to yield the total scattering structure function, $S(Q)$.  This is 
Fourier transformedaccording to 
\begin{equation}
  G(r)= \frac{2}{\pi }\int_{0}^{\infty} Q [S(Q)-1] \sin (Qr)\> dQ.
\end{equation}
Data collection and analysis procedures have been described
elsewhere.~\cite{billi;prb93} The reduced structure factor $F(Q)=Q[S(Q)-1]$
is shown in Figure \ref{fig;soq}. 

\section{Modelling and Results}

The Rietveld refinements were carried out using the GSAS Rietveld
code.\cite{larso;unpub87} Modelling of PDF was carried out using a 
least-squares full-profile PDF fitting procedure.  This is is exactly 
analogous to the Rietveld method except that the PDF is fit (in real-space) 
rather than the reciprocal-space data.  When the PDF is fit the 
short-range order is obtained directly.~\cite{billi;b;pom98} The program 
we use is called PDFFIT. It is described in detail elsewhere and is 
available on request.~\cite{prbi99} The structural inputs for the 
program are atomic positions, occupancies and thermal factors.  

The results are shown in Table~\ref{tab;struc}.  We
chose the convention used in Ref.~\onlinecite{rodri;prb98} of putting Mn on
the (0,$\frac{1}{2},0$) position.
\begin{table}[!tb]
  \centering
  \begin{tabular}{crrr}
                   &  Rietveld     &  Refinement A & Refinement B \\
  \hline
  a                &  5.542(1)     &   5.5422(7)   &  5.557(1)   \\
  b                &  5.732(1)     &   5.7437(8)   &  5.774(1)   \\
  c                &  7.6832(2)    &   7.690(1)    &  7.712(2)   \\
                   &               &               &             \\
  x(La)            &  -0.0068(3)   &  -0.0073(2)   & -0.0068(2)  \\
  y(La)            &   0.0501(3)   &   0.0488(2)   &  0.0504(1)  \\
  \u2(La)          &   0.0022(4)   &   0.00199(4)  &  0.00177(4) \\
                   &               &               &             \\
  \u2(Mn)          &   0.0011(6)   &   0.00067(7)  &  0.00071(7) \\
                   &               &               &             \\
  x(O1)            &   0.0746(4)   &   0.0729(3)   &  0.0734(2)  \\
  y(O1)            &   0.4873(4)   &   0.4857(3)   &  0.4800(2)  \\
  \u2(O1)          &   0.0031(5)   &   0.00233(7)  &  0.00178(5) \\
                   &               &               &             \\
  x(O2)            &   0.7243(3)   &   0.7247(3)   &  0.7232(2)  \\
  y(O2)            &   0.3040(3)   &   0.3068(3)   &  0.3060(1)  \\
  z(O2)            &   0.0390(2)   &   0.0388(3)   &  0.0369(2)  \\
  \u2(O2)          &   0.0030(4)   &   0.00378(5)  &  0.00439(4) \\
                   &               &               &             \\
  R$_{wp}$         &  12.1         &  16.2         &  9.1        \\
  \end{tabular}
  \caption{Structural data of LaMnO$_{3}$ ($Pbnm$) from the 
           Rietveld refinement and PDF refinements 
           A and B. La and O1 are on (x,y,$\frac{1}{4}$),
           Mn is on (0,$\frac{1}{2},0$) and O2 is on (x,y,z). The 
           units for the lattice parameters are \AA\ and for values
           of \u2 \AA$^{2}$. The numbers in parentheses are the estimated 
	   standard deviation on the last digit.}
  \label{tab;struc}
\end{table}
Two PDF refinements are reported labelled A and B.  The difference is the
range of $r$ over which the fit was made: A was made over a range
${\rm 1.5~\AA<{\mit r}<15.5~\AA }$; B over a range
${\rm 1.5~\AA<{\mit r}<3.5~\AA }$. Both refinements were
constrained to have the symmetry of the $Pbnm$ space group.
In addition, PDF-refinements were carried out where the space group
symmetry was relaxed.  However, the results of 
these refinements essentially reproduced
those within the $Pbnm$ space group and the results are not reported here.

\begin{figure}[!tb]
  \centering
  \includegraphics[angle=0,width=2.8in]{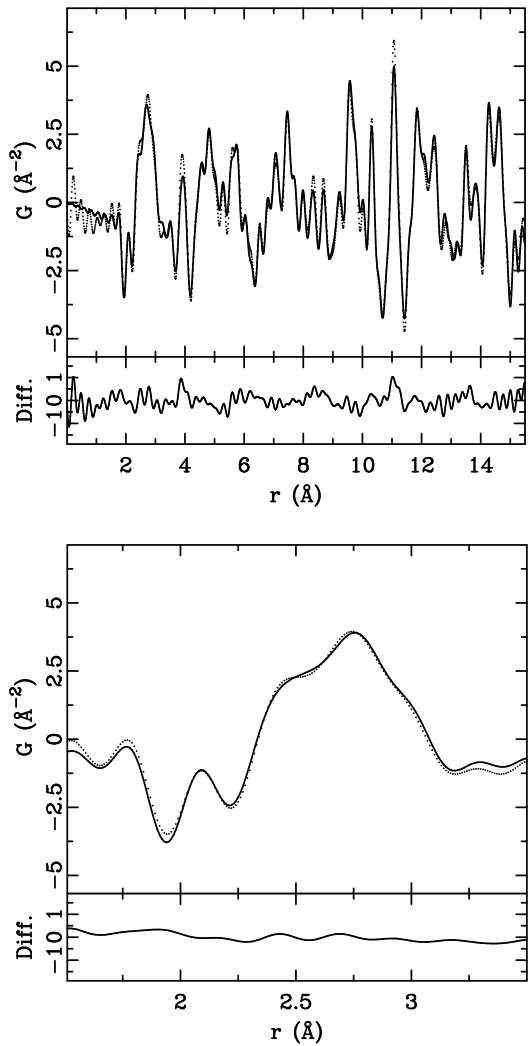}
  \caption{Experimental PDF (circles) and model PDF (solid line) 
           for refinements A (top) and B (bottom). Difference curves are 
           plotted below the data.}
  \label{fig;pdf}
\end{figure}

Of particular interest are the resulting MnO bond lengths in the MnO$_{6}$ 
octahedra, listed in Table \ref{tab;bond}. The R-values given in 
Table \ref{tab;bond} are calculated over the same interval 1.5 to 3.5~\AA\ 
so they can be directly compared with each other. The observed and
calculated PDFs for runs A and B are shown in Figure~\ref{fig;pdf}.
\begin{table}[!tb]
  \centering
  \begin{tabular}{crrrr}
  Run      &  Mn-O ($s$)   &  Mn-O ($m$)    &  Mn-O ($l$) & R$_{wp}$ \\
  \hline
  Rietveld &  1.9200(3)    &  1.9662(4)     &  2.1609(4)  &    -     \\
  A        &  1.910(3)     &  1.9662(8)     &  2.178(3)   & 13.1     \\
  B        &  1.924(2)     &  1.9742(9)     &  2.177(2)   &  9.1     \\
  \end{tabular}
  \caption{Mn-O bond lengths in units of \AA . For details see text.}
  \label{tab;bond}
\end{table}

In addition, we have determined the Debye Temperature of the Mn and Oxygen
ions from the refinements.  The data were collected at 20~K. Assuming this
is close enough to 0~K we use the expression
\begin{equation}
  \Theta_{D} = \frac{3 h^{2}}{16\pi^{2} m k_{b} \langle u^{2} \rangle}
  \label{eq;thetad}
\end{equation}
to determine the $\Theta_{D}$ from the refined thermal displacements,
\u2 of Mn and O atoms, here $m$ is the mass of the corresponding atom
and $k_{b}$ Boltzmann's constant.  
Thermal factors obtained from PDF refinements are often more accurate than
Rietveld thermal factors because of the
wider range of $Q$ over which data are analyzed~\cite{billi;b;lsfd98}.
For example, in this study the PDF's were obtained from data collected
up to $Q_{max}= 27$~\AA$^{-1}$.  This is almost double the $Q$-range
used ($Q_{max}= 15.7$~\AA$^{-1}$) in the Rietveld refinement of the same
data.  The Rietveld refinement was confined to a lower $Q$-range due to
Bragg-peak overlap in the high-$Q$ region.  We have previously shown that  
the PDF can give accurate absolute values of 
$\Theta_{D}$~\cite{jeong;jpc99}.
The values we obtain are $\Theta_{D}$(Mn)$= 1000 \pm 100$~K, 
$\Theta_{D}$(O1)$= 980 \pm 30$~K and $\Theta_{D}$(O2)$= 601 \pm 8$~K.


\section{Discussion}

The average crystallographic structure suggests that the Jahn-Teller distorted
octahedra in LaMnO$_3$ contain two short ($s$) bonds (1.9200~\AA), two less 
short ($m$) bonds (1.9662~\AA) and two long ($l$) bonds (2.1609~\AA).  The PDF
peaks corresponding to these Mn-O bonds can be seen in Fig.~\ref{fig;pdf} at 
around $r=2$~\AA\ as negative peaks.\cite{footnote;lmound} A double-peak 
structure is clearly resolved reflecting the high resolution of the PDF 
measurement.  

The motivation for this study was to determine whether the real, local, 
JT distorted octahedra had 4-$s$ and 2-$l$ bonds (pure $Q_3$ 
distortion\cite{sturg;b;ssp67}), which one would expect for an isolated 
octahedron, or 2-$s$, 2-$m$ and 2-$l$ bonds (some $Q_2$ component) as 
suggested by the average structure. The JT distorted octahedra could be 
locally $Q_3$ but appear further distorted in the average structure if 
there was some orbital disorder (for example, some of the long bonds 
orienting parallel to the $c$-axis). These two scenarios can be distinguished 
in a joint Rietveld/PDF study where the average and local structures are 
determined from the same data-set.

When the PDF is fit over a wide $r$-range (run A) the Mn-O bond lengths 
are similar to the Rietveld values, suggesting that the local bond-length 
distribution matches that of the average structure.  However, it is 
possible that even by $r=15$~\AA\ the effects of possible orbital disorder 
will bias the results of the PDF refinement towards the average values.  
To check this, we carried out run B which fits only over the range to 
3.5~\AA.  This PDF-range contains only the
MnO$_6$ octahedra themselves and does not depend on how they are oriented
in space.  It is clear from Table~\ref{tab;bond} that the refinement
still prefers two short, two less-short and two long bonds.  As a final check
we carried out refinements where the $Pbnm$ space-group symmetry was
relaxed so that all Mn  and O ion positions can vary independently.  This 
allows up to six different Mn-O bond-lengths to refine within one octahedron.  
The refinements again resulted in the bond lengths grouping into 2 short, 
2 less short and 2 long.
It is clear that the PDF peak corresponding to the short octahedral bonds
is broader than can be explained by a single Mn-O bond-length.  This is
strong evidence that the local Jahn-Teller distortion is not a pure 
stretch of the $d_{3z^2-r^2}$ orbitals but there is a small amount of
$d_{x^2-y^2}$ character mixed in and there is some $Q_2$ character to
the static distortion.
\begin{figure}[!tb]
  \centering
  \includegraphics[angle=0,width=2.8in]{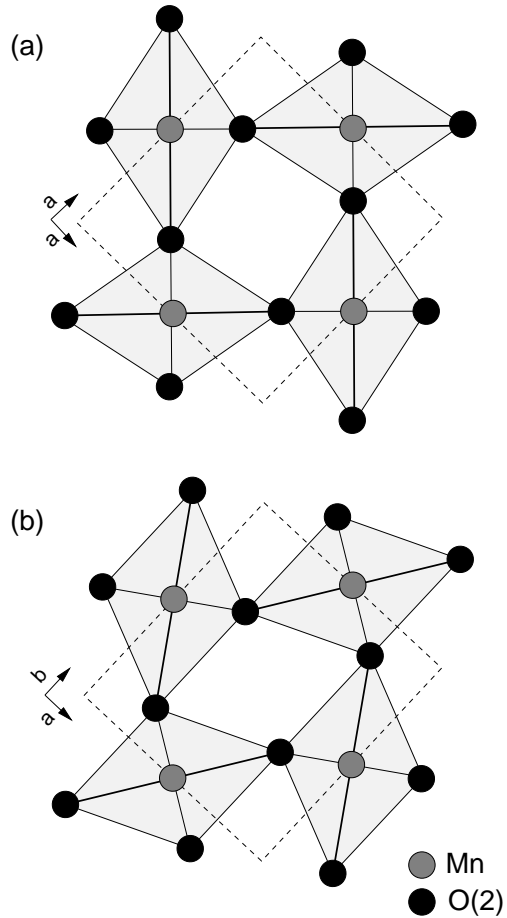}
  \caption{Schematic drawing of the arrangement of MnO$_{6}$ octahedra
           projected down the $c$-axis. The long distance in the octahedra
           corresponds to the $d_{3z^{2}-r{2}}$ orbital, the short 
           distance to $d_{x^{2}-y^{2}}$. (a) the octahedral rotations
	   have been set to zero resulting in a square based unit cell 
	   ($a=b$).
           (b) Finite rotations result in the orthorhombic structure ($a \ne
	   b$) that is observed.}
  \label{fig;oct}
\end{figure}

This can be explained by the fact that the {\it average} structure is 
orthorhombic and so the octahedra are sitting in an orthorhombic crystal field.
It is helpful to understand the origin of the long-range orthorhombicity
of the structure itself. This comes about because of the rotations 
(about $\langle 111\rangle$ directions) of the corner-shared octahedra.  
This does not in itself result in an orthorhombic distortion; however, if 
the octahedra themselves are elongated and the orbitals ordered, as in 
this case, an orthorhombic distortion does result.  
The basal-plane distortion ($2(a-b)/(a+b)$ where $a$ refers to the axis 
of the orthorhombic cell) is explained straightforwardly by reference to 
Fig.~\ref{fig;oct}(b). In this figure the octahedra are shown elongated 
and it is clear how the rotations give rise to the basal plane distortion.  

The manganese-manganese separation along the perpendicular direction is 
also different to the basal plane, but this is due to a different reason.
With the pattern of orbital ordering in this O$^\prime$ structure, 
all of the long Mn-O bonds lie in the basal plane and all the Mn-O bonds 
along the $c$-direction are short. The separation of manganese ions along
the $c$ direction is therefore shorter than in the basal plane
and $c/\sqrt{2}\le a \le b$. This would still hold if the shapes
of the octahedra themselves were tetragonal (pure $Q_3$ distortion)
and comes simply from the fact that the long bonds all lie in the basal 
plane.  Thus, the structure could accommodate having tetragonally ($Q_3$) 
distorted octahedra within the orthorhombic unit cell by choosing the 
appropriate $c$-axis lattice parameter to make the Mn-O(2) bond along 
the $c$-direction the same as the Mn-O(1) (in-plane) short bond.  
In fact this does not happen.  

What is clear from the local Mn-O bond-lengths is that there is a large
$Q_3$ distortion with a small $Q_2$ distortion superimposed.  The large
$Q_3$ distortion breaks the degeneracy of the $e_g$ orbitals and separates
them widely in energy ($\sim 0.5$~eV).\cite{milli;prl95}  The additional
small $Q_2$ distortion comes about by mixing the pure 
$d_{3z^2-r^2}$ and $d_{x^2-y^2}$ but is presumably a small response
of the local octahedra to the orthorhombic crystal field, even though
this long-range orthorhombicity results from the large local $Q_3$ distorted
octahedra (and their rotations).


\section{Conclusions}

We have fit a high-resolution atomic pair distribution function obtained
from powder neutron diffraction data to determine the local Jahn-Teller
distortion in stoichiometric LaMnO$_3$.  We observe a small but significant
difference in the length of the in-plane (1.924(2)~\AA) and 
out-of plane (1.9742(9)~\AA) short bonds, and a well separated long bond 
(2.177(2)~\AA).  This implies that there is some mixing of the 
$d_{3z^2-r^2}$  and $d_{x^2-y^2}$ levels and the occupied $e_g$ state 
is not pure $d_{3z^2-r^2}$ in character.  This is in agreement with the 
result from crystallography; however, it is important to determine this 
directly from the local structure as we report here since any
orbital disorder (for example, due to small non-stoichiometries) would
affect the crystal structure but not the local structure.   
Finally, we report estimates of the Debye temperature of Mn and O ions
in this compound.


\acknowledgements
We would like to acknowledge stimulating discussions with S. D. Mahanti,
P. Radaelli and T. A. Kaplan.
This work was supported by the NSF through grant DMR-9700966 at MSU and 
the DOE through contract W-7405-ENG-36 at LANL.  The IPNS is funded by 
the U.S. Department of Energy under Contract W-31-109-Eng-38.




\begin{thebibliography}{10}
\bibitem{footnote;lmound}
The PDF peak originating from the pair of atoms $i$ and $j$ is weighted by the
  product $b_ib_j$ where $b_i$ is the scattering length of atom $i$. Since the
  neutron scattering length of Mn is negative, the Mn-O peaks in the PDF are
  negative.
\end{thebibliography}
\end{document}